\documentclass[preprint,12pt]{elsarticle}

\usepackage{amssymb}
\usepackage{multirow}
\usepackage{graphicx}
\usepackage{array, booktabs, longtable}	%for long long table
\usepackage{lineno,hyperref}
\usepackage{multirow}
\usepackage{algorithm}
\usepackage[noend]{algpseudocode}
\journal{journal for review}
\begin{document}

\begin{frontmatter}



\title{Recognition of Ischaemia and Infection in Diabetic Foot Ulcers: Dataset and Techniques}

\author[1]{Manu Goyal}
\author[2]{Neil D. Reeves}
\author[3]{Satyan Rajbhandari}
\author[4]{Naseer Ahmad}
\author[5]{Chuan Wang}
\author[1]{Moi Hoon Yap \corref{cor1}}
\cortext[cor1]{Corresponding author: 
	Tel.: +44 161 247 1503;  
}
\ead{M.Yap@mmu.ac.uk}

\address[1]{Centre for Advanced Computational Sciences, Manchester Metropolitan University, M1 5GD, Manchester, UK}
\address[2]{Research Centre for Musculoskeletal Science \& Sports Medicine, Manchester Metropolitan University, M1 5GD, Manchester, UK.}
\address[3]{Lancashire Teaching Hospital, PR2 9HT, Preston, UK.}
\address[4]{University of Manchester and Manchester Royal Infirmary, M13 9WL, Manchester, UK.}
\address[5]{Department of Endocrinology, Sun Yat-sen Memorial Hospital, Sun Yat-sen University, Guangzhou 510120, P.R. China.}
\address{}

\begin{abstract}
		Recognition and analysis of Diabetic Foot Ulcers (DFU) using computerized methods is an emerging research area with the evolution of image-based machine learning algorithms. Existing research using visual computerized methods mainly focuses on recognition, detection, and segmentation of the visual appearance of the DFU as well as tissue classification. According to DFU medical classification systems, the presence of infection (bacteria in the wound) and ischaemia (inadequate blood supply) has important clinical implications for DFU assessment, which are used to predict the risk of amputation. In this work, we propose a new dataset and computer vision techniques to identify the presence of infection and ischaemia in DFU. This is the first time a DFU dataset with ground truth labels of ischaemia and infection cases is introduced for research purposes. For the handcrafted machine learning approach, we propose a new feature descriptor, namely the Superpixel Color Descriptor. Then we use the Ensemble Convolutional Neural Network (CNN) model for more effective recognition of ischaemia and infection. We propose to use a natural data-augmentation method, which identifies the region of interest on foot images and focuses on finding the salient features existing in this area. Finally, we evaluate the performance of our proposed techniques on binary classification, i.e. ischaemia versus non-ischaemia and infection versus non-infection. Overall, our method performed better in the classification of ischaemia than infection. We found that our proposed Ensemble CNN deep learning algorithms performed better for both classification tasks as compared to handcrafted machine learning algorithms, with 90\% accuracy in ischaemia classification and 73\% in infection classification. 

\end{abstract}




\end{frontmatter}

%\linenumbers

\section{Introduction}
    Diabetic Foot Ulcers (DFUs) are a major complication of diabetes which can lead to amputation of the foot or limb. Treatment of Diabetic foot ulcers is a global major health care problem resulting in high care costs and mortality rate. Recognition of infection and ischaemia is very important to determine factors that predict the healing progress of DFU and the risk of amputation. Ischaemia, the lack of blood circulation, develops due to chronic complications of diabetes. This can result in gangrene of the diabetic foot ulcer, which may require amputation of the part of the foot or leg if not recognised and treated early. Detailed knowledge of the vascular anatomy of the leg, and particularly ischaemia enables medical experts make better decisions in estimating the possibility of DFU healing, given the existing blood supply \cite{santilli1999chronic}. In previous studies, it is estimated that patients with critical ischaemia have a three-year limb loss rate of about 40\% \cite{albers1992assessment}. Patients with an active DFU and particularly those with ischaemia or gangrene should be checked for the presence of infection. Approximately, 56\% of DFU become infected and 20\% of DFU infections lead to amputation of a foot or limb \cite{prompers2007high, lipsky20122012, lavery2003diabetic}. In one recent study, 785 million patients with diabetes in the US between 2007 and 2013 suggested that DFU and associated infections constitute a powerful risk factor for emergency department visits and hospital admission \cite{skrepnek2017health}.

	There are a number of DFU classification systems such as  Wagner, University of Texas, and SINBAD Classification systems, which include information on the site of DFU, area, depth, presence of neuropathy, presence of ischaemia, and infection \cite{wagner1987diabetic, lavery1996classification, ince2008use}. SINBAD stands for S (Site), I (Ischaemia), N (Neuropathy), B (Bacterial infection), A (Area), D (Depth). This paper focuses on ischaemia and infection, which are defined as follow:
	
\begin{enumerate}
	\item Ischaemia: Inadequate blood supply that could affect DFU healing. Ischaemia is diagnosed by palpating foot pulses and measuring blood pressure in the foot and toes. The visual appearance of ischaemia might be indicated by the presence of poor reperfusion to the foot, or black gangrenous toes (tissues death to part of the foot). From a computer vision perspective, these might be important hints of the presence of ischaemia in the DFU.  
	
	\item Bacterial Infection: Infection is defined as bacterial soft tissue or bone infection in the DFU, which is based on the presence of at least two classic findings of inflammation or purulence. It is very hard to determine the presence of diabetic foot infections from DFU images, but increased redness in and around ulcer and coloured purulent could provide indications. In the medical system, blood testing is performed as the gold standard diagnostic test. Also, in the present dataset, the images were captured after the debridement of necrotic and devitalized tissues which removes an important indication of the presence of infection in DFU.   
	
\end{enumerate}

In related work, Netten et al. \cite{van2017validity} find that clinicians achieved low validity and reliability for remote assessment of DFU in foot images. Hence, it is clear that analysing these conditions from images is a difficult task for clinicians. In various image recognition applications, such as medical imaging and natural language processing tasks, machine learning algorithms performed better than skilled humans including clinicians \cite{gulshan2016development, esteva2017dermatologist, krizhevsky2012imagenet}. 

The previous state-of-the-art image-based computer-aided diagnosis of DFU is composed of multiple stages, including image pre-processing, image segmentation, feature extraction, and classification. Veredas et al. \cite{veredas2009binary} proposed the use of color and texture features from the segmented area and multi-layer neural network to perform the tissue classification to distinguish between healing-tissue and skin for healing prediction. Wannous et al. \cite{wannous2010enhanced} performed tissue classification from color and texture region descriptors on a 3-D model for the wound. Wang et al. \cite{wang2016area} used a cascaded two-stage classifier to determine the DFU boundaries for area determination of DFU. Major progress in the field of image-based machine learning, especially deep learning algorithms, allows the extensive use of medical imaging data with end-to-end models to provide better diagnosis, treatment, and prediction of diseases \cite{yap2018breast, ahmad2018semantic}. Deep learning models for DFU, predominantly led by works from our laboratory have achieved high accuracy in the recognition of DFUs with machine learning algorithms \cite{goyal2018dfunet, goyal2017fully, goyal2018robust, wang2015unified}.

The major issues and challenges involved with the assessment of DFU using machine learning methods from foot images are as follows: 1) a major time-burden involved in data collection and expert labelling of the DFU images; 2) high inter-class similarity and intra-class variations are dependent upon the different classification of DFU; 3) non-standardization of the DFU dataset, such as distance of the camera from the foot, orientation of the image and lighting conditions; 4) lack of meta-data, such as patient ethnicity, age, sex and foot size. 

Accurate diagnosis of ischaemia and infection requires establishing a good clinical history, physical examination, blood tests, bacteriological study and Doppler study of leg blood vessels. These tests and resources are not always available to clinicians across the world and hence the need for a solution to inform diagnosis, such as the one we proposed in this paper. Experts working in the field of diabetic foot ulceration have good experience of predicting the presence of underlying ischaemia or infection simply by looking at the ulcer. We aim to replicate that in machine learning. To increase the reliability of the annotation, two experts predict the presence of ischaemia and infection from DFU images. Due to high risks of infection and ischaemia in DFU leading to patient's hospital admission, and amputation \cite{mills2014society}, recognition of infection and ischaemia in DFU with cost-effective machine learning methods is a very important step towards the development of complete computerized DFU assessment system for remote monitoring in the future.

\section{DFU Dataset and Expert Labelling}\label{sec:7.2}

For binary classification of ischaemia and infection in DFU, we introduce a dataset of 1459 images of patient's foot with DFU over the previous five years at the Lancashire Teaching Hospitals, obtaining ethical approval from all relevant bodies and patient’s written informed consent. Approval was obtained from the NHS Research Ethics Committee to use these images for this research. These DFU images were captured with different cameras (Kodak DX4530, Nikon D3300, and Nikon COOLPIX P100). The current dataset we received with the ethical approval from NHS did not contain any records or meta-data about these conditions or any medical classification.

\begin{figure*}[!t]
	\centering
	\small
	\begin{tabular}{ccc}
		\includegraphics[width=6cm,height=6cm]{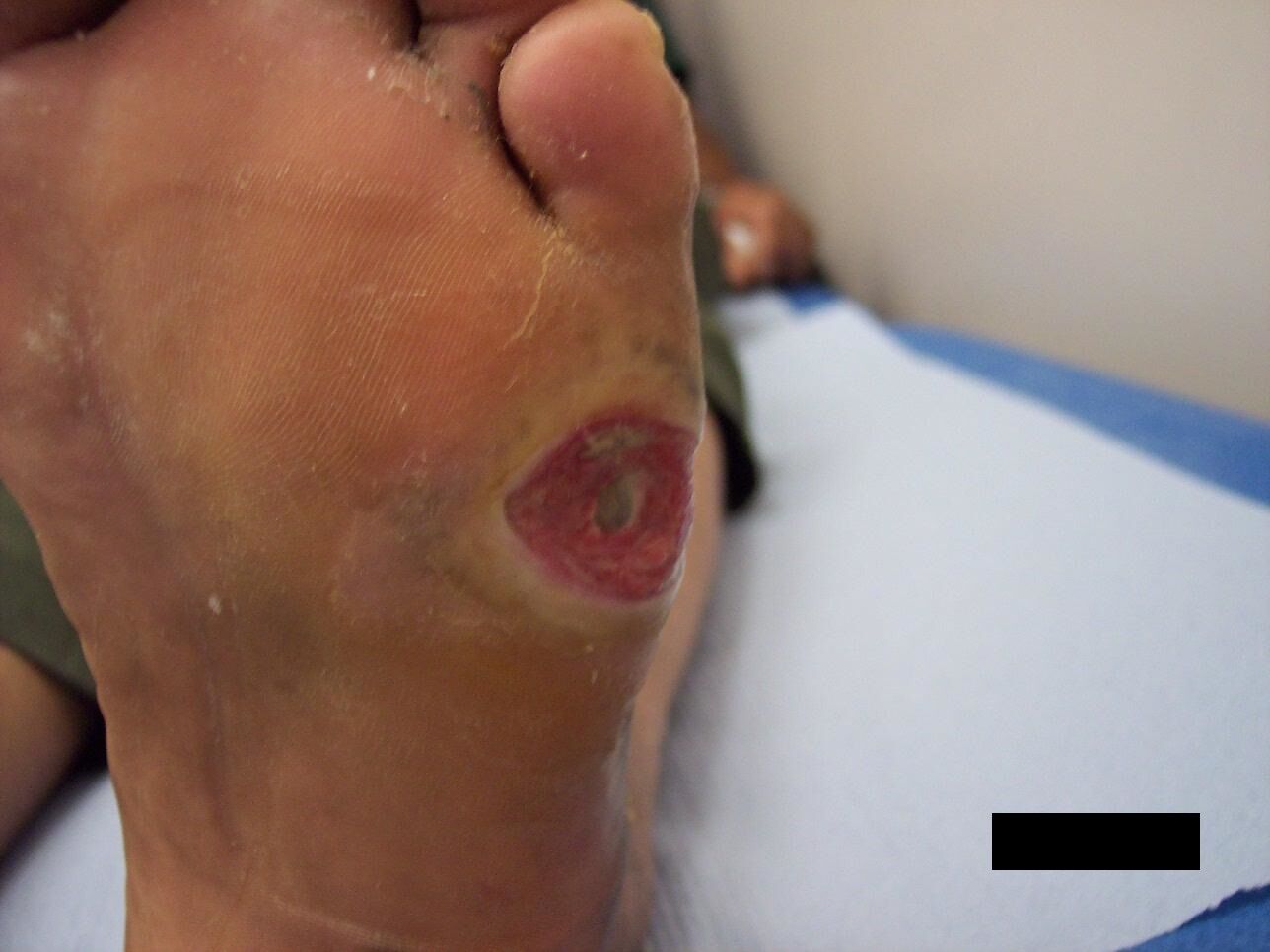} &
		\includegraphics[width=6cm,height=6cm]{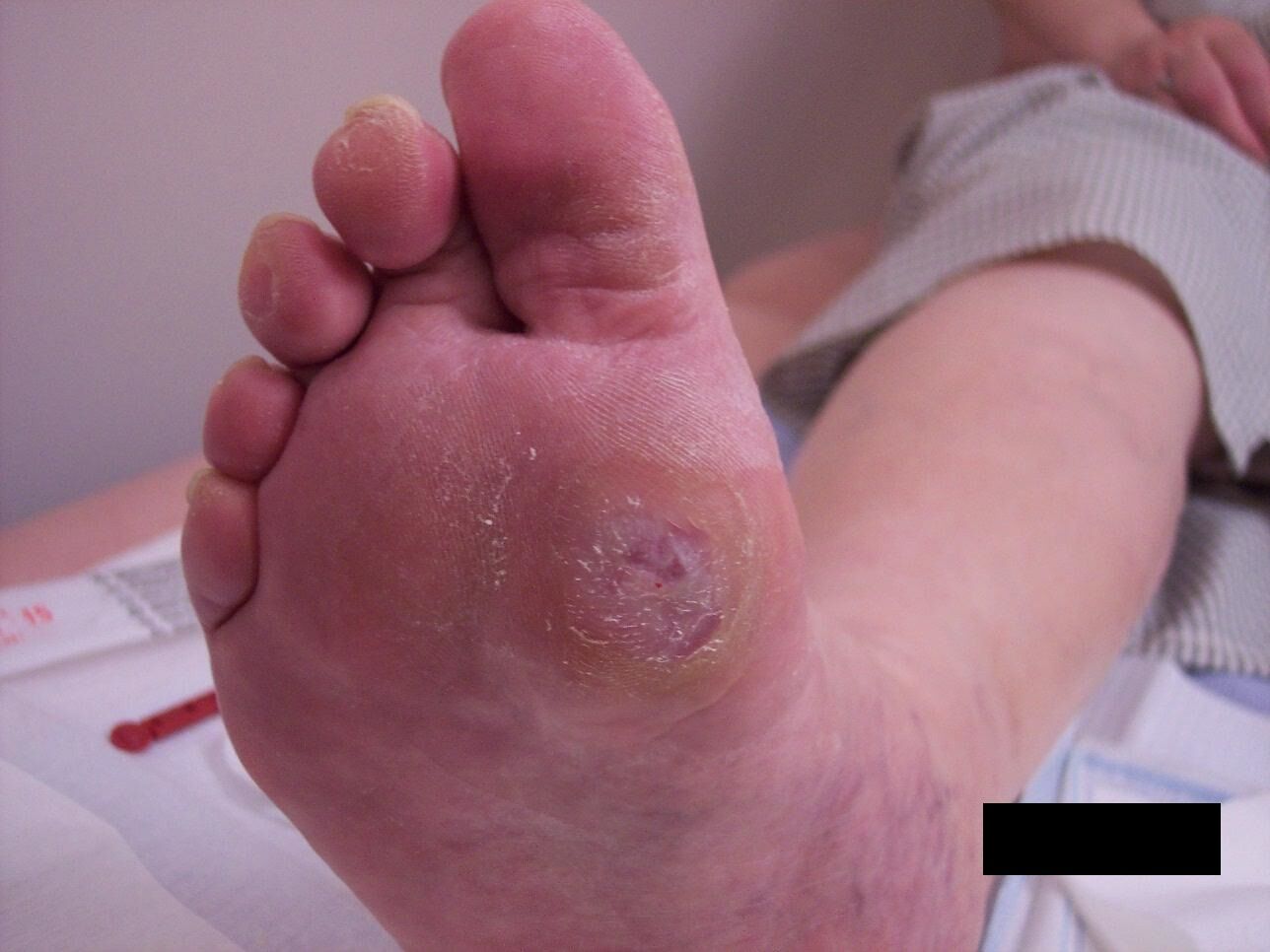}\\
		(a) Infection & (b) No Infection \\
		\includegraphics[width=6cm,height=6cm]{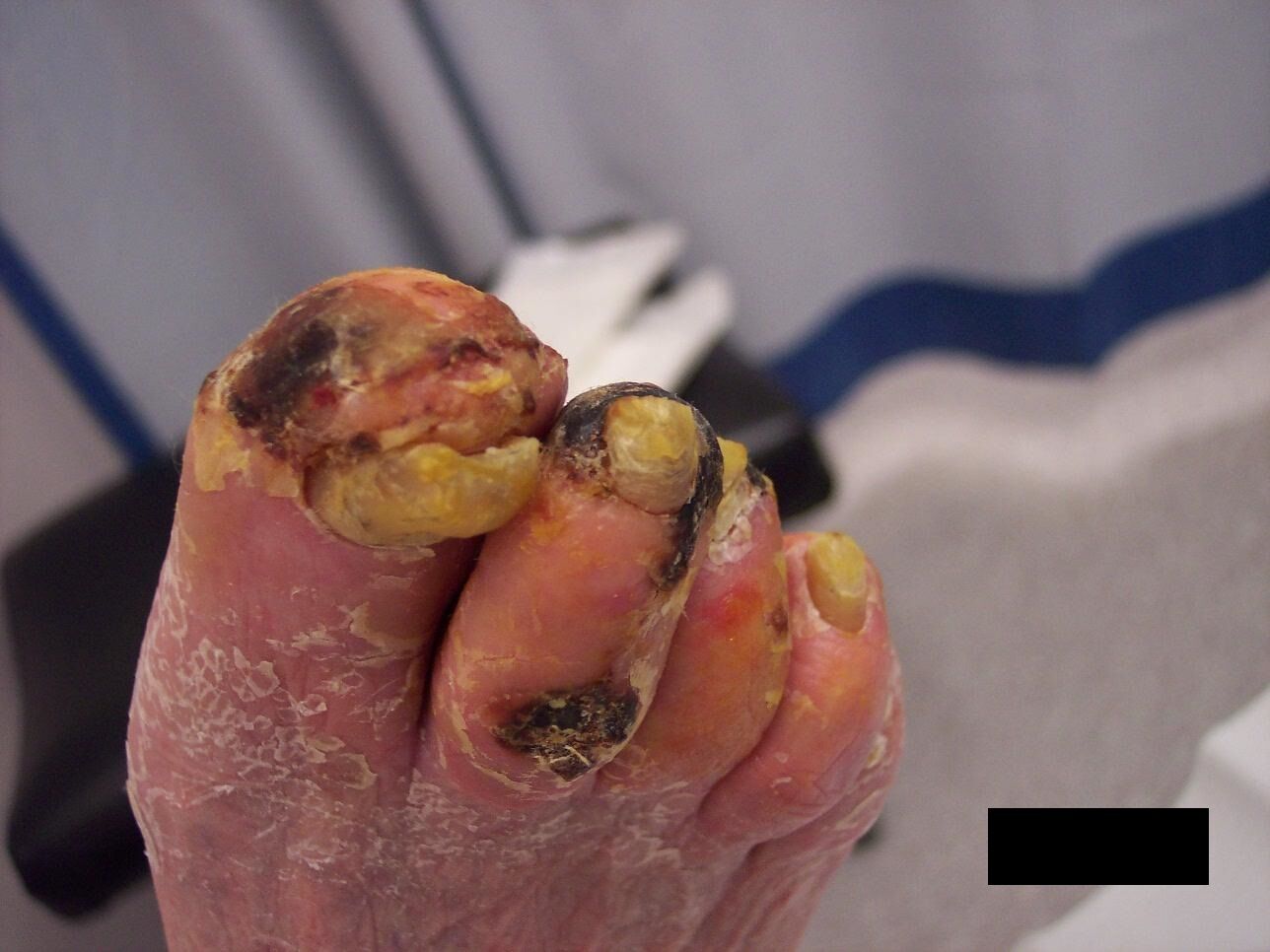}& 
		\includegraphics[width=6cm,height=6cm]{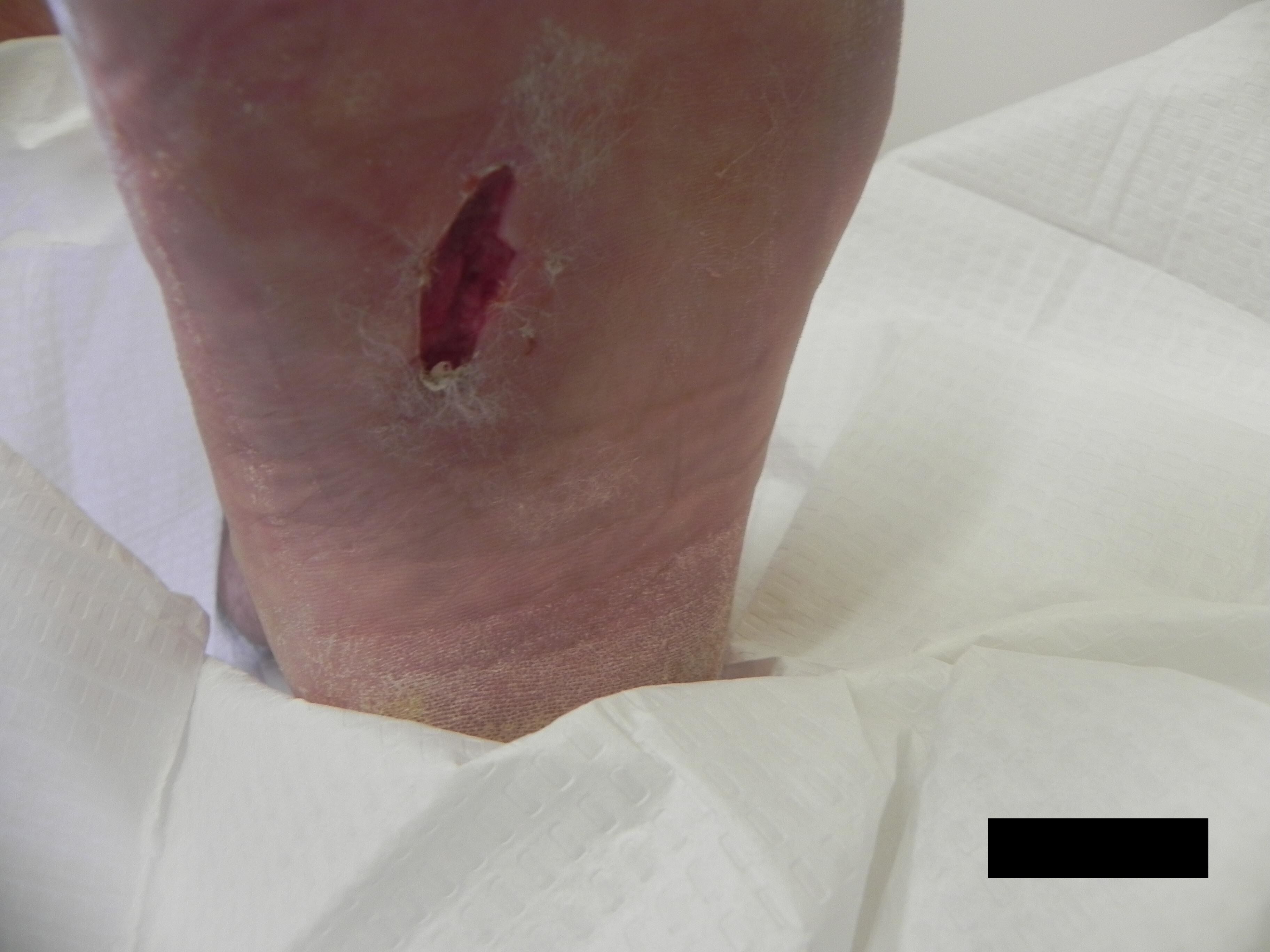}\\  
		(c) Ischaemia & (d) No Ischaemia \\
	\end{tabular}     
	
	\caption{Examples of foot images with DFU used for binary expert annotations for infection and ischaemia.}
	\label{fig:Examples}
\end{figure*}

\begin{figure}
	\centering
	\includegraphics[scale=0.60]{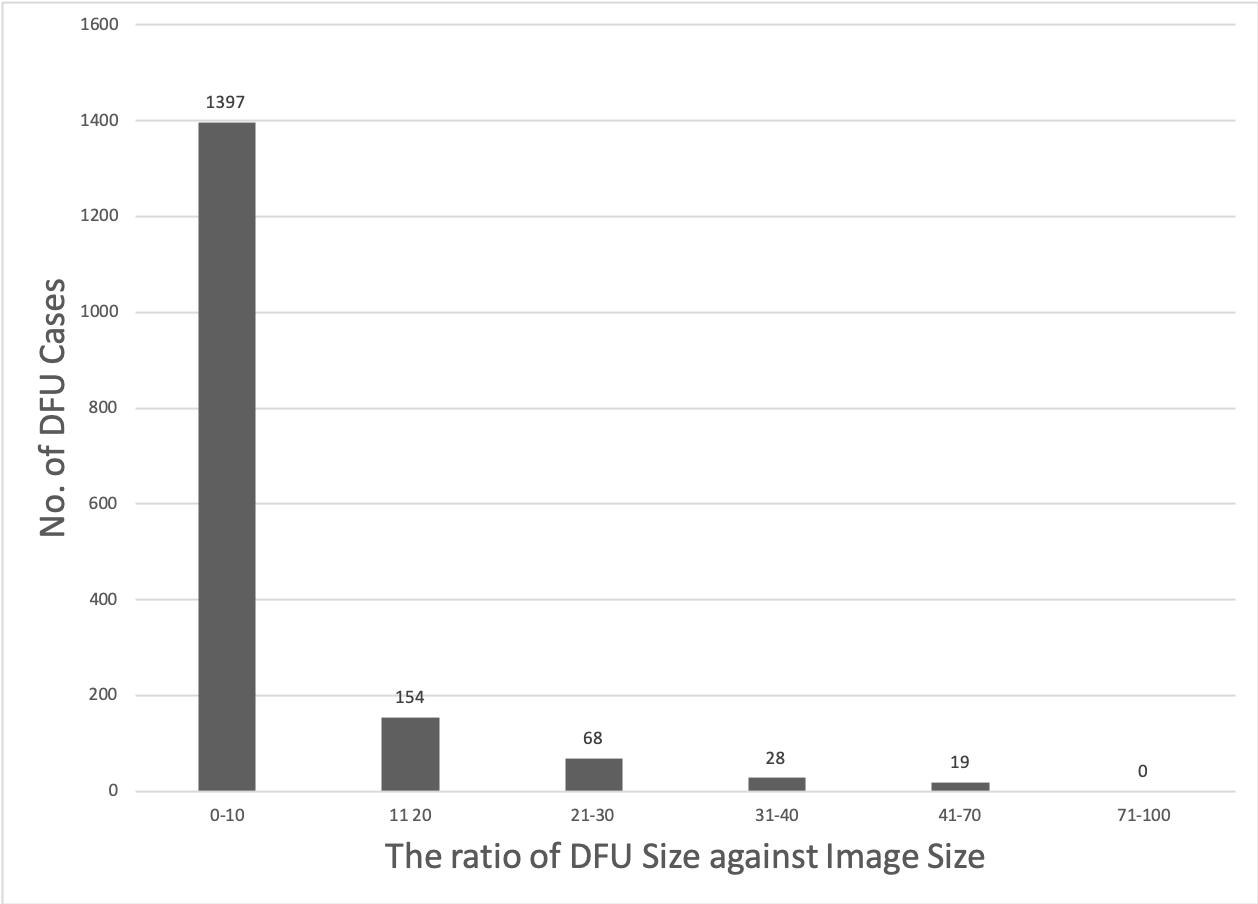}
	\caption{The number of DFU cases according to the area of DFU in full foot image of the DFU dataset.}
	\label{fig:AreaDFUs}
\end{figure}

Since there is no clinical meta-data regarding this DFU dataset, the experiment is performed on the images with handcrafted traditional machine learning and deep learning. This is the first time, recognition of ischaemia and infection in DFU is performed based on images, hence, there is no publicly available dataset. Here, we introduce the first DFU dataset with ground truth labels of ischaemia and infection cases. Expert labelling of each DFU according to the different conditions present in DFU according to the popular medical classification system on this DFU dataset is particularly important for this task. The ground truth was produced by two healthcare professionals (consultant physicians with specialisation in the diabetic foot) on the visual inspection of DFU images. Where there was disagreement for the ground truth, the final decision was made by the more senior physician. These ground truths are used for the binary classification of infection and ischaemia of DFU. A few examples of foot images with DFU used for binary expert annotation are shown in Fig. \ref{fig:Examples}. The complete number of cases of expert annotation of each condition is detailed in Table \ref{tab:sinbad2}. The dataset, alongside its ground truth labels, will be made available upon acceptance of this article.

\section{Methodology}
This section describes our proposed techniques for the recognition of ischaemia and infection of the DFU diagnosis system. The preparation of a balanced dataset, handcrafted features, and machine learning methods (handcrafted machine learning and deep learning approaches) used for binary classification of ischaemia and infection are detailed in this section. 

\subsection{Natural Data-Augmentation Technique based on Deep Learning Algorithm}
This section describes our proposed data augmentation method, called Natural Data-augmentation, which is based on deep DFU localization algorithm (Faster R-CNN).

\begin{figure*}[!t]
	\centering
	\small
	\begin{tabular}{ccccc}
		\includegraphics[width=3cm,height=3cm]{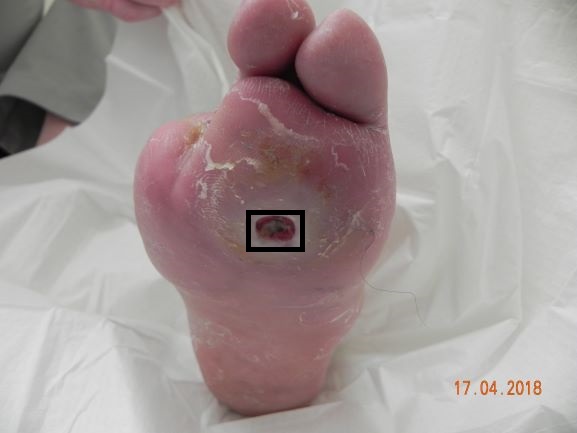} &
		\includegraphics[width=3cm,height=3cm]{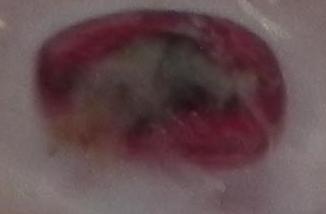}&
		\includegraphics[width=3cm,height=3cm]{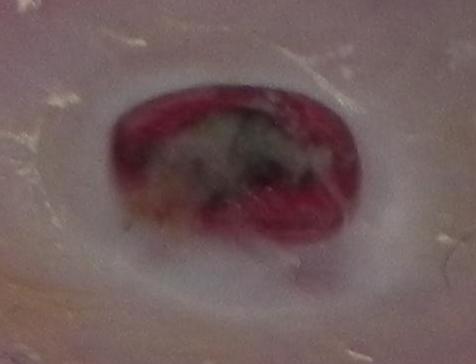}& 
		\includegraphics[width=3cm,height=3cm]{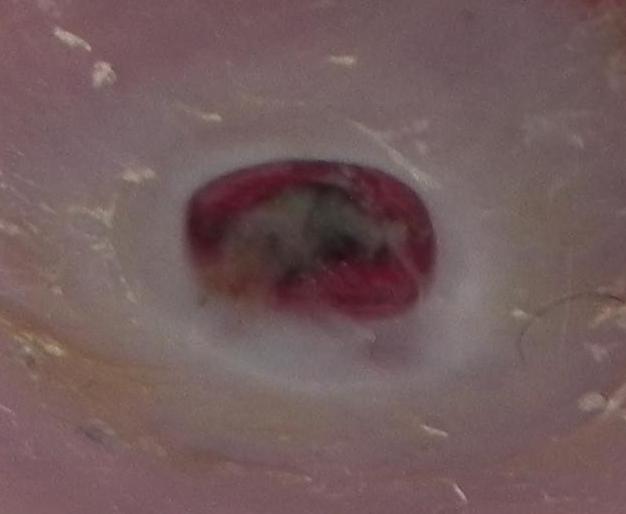}\\  
		\\
		\includegraphics[width=3cm,height=3cm]{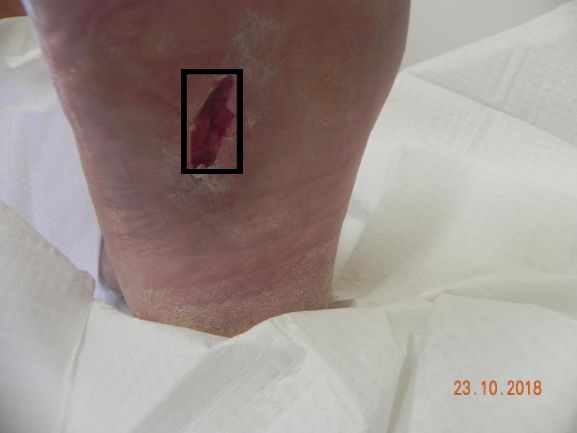} &
		\includegraphics[width=3cm,height=3cm]{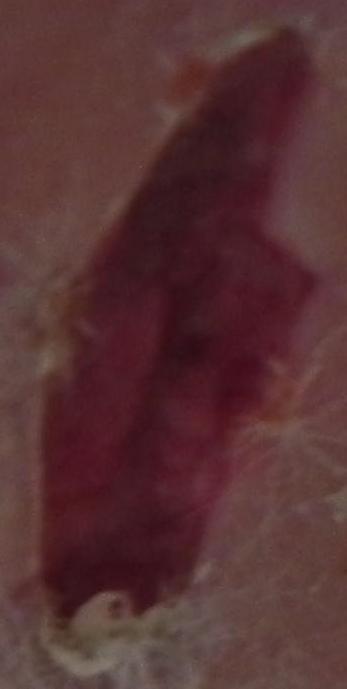}&
		\includegraphics[width=3cm,height=3cm]{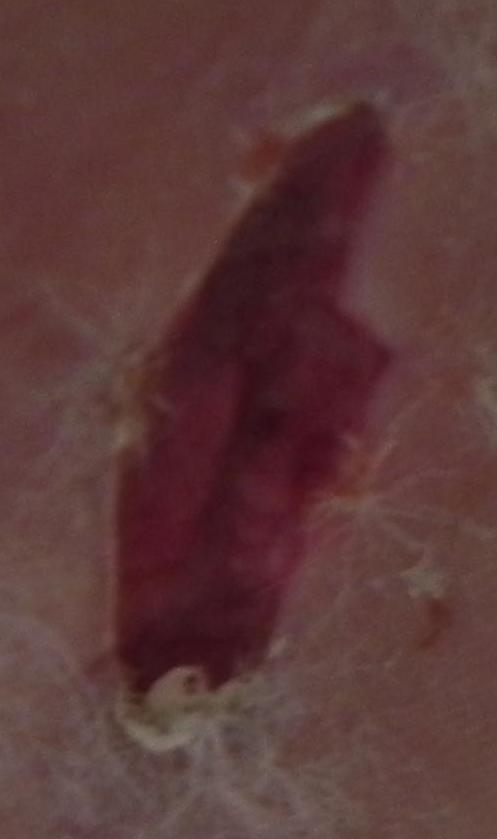}& 
		\includegraphics[width=3cm,height=3cm]{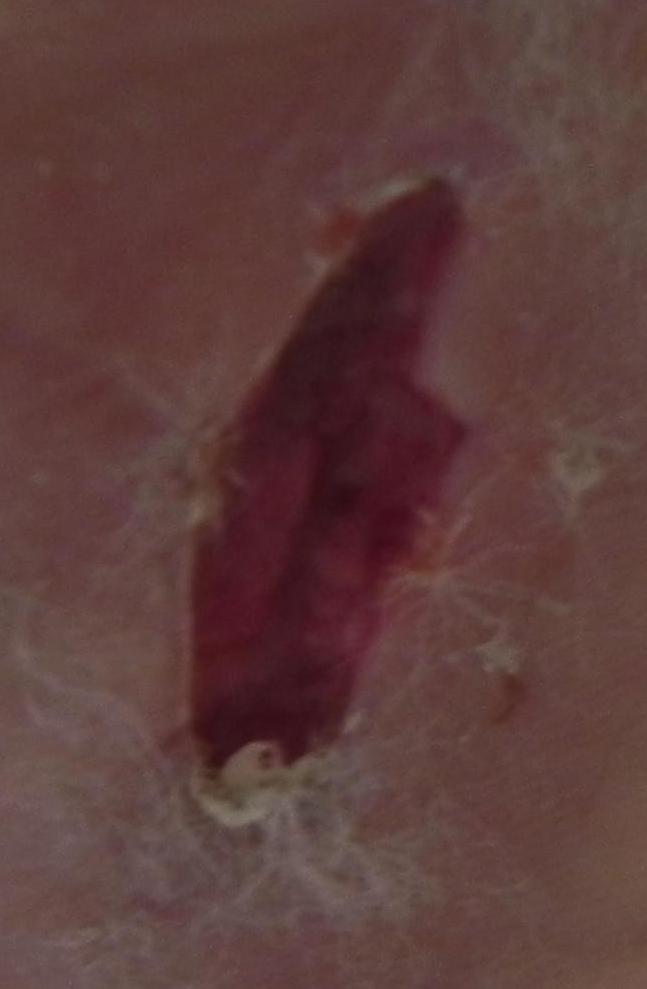}&    
		\\
		\includegraphics[width=3cm,height=3cm]{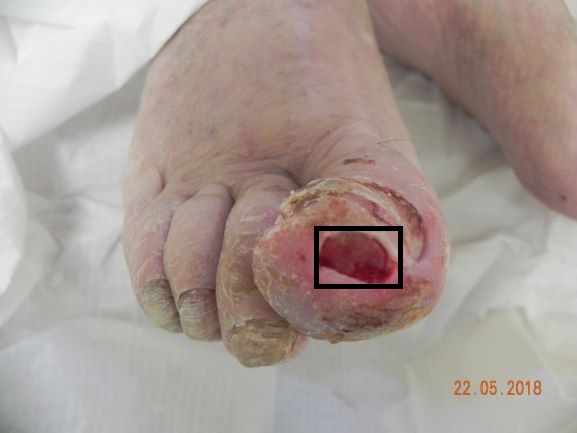} &
		\includegraphics[width=3cm,height=3cm]{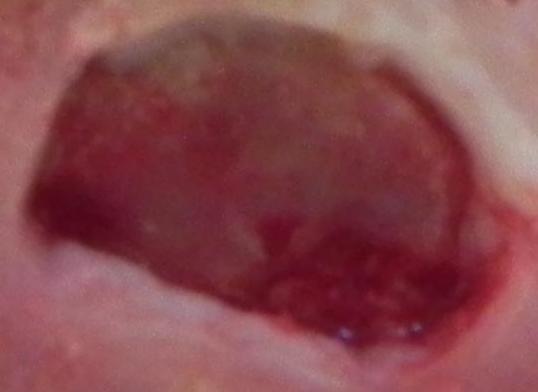}&
		\includegraphics[width=3cm,height=3cm]{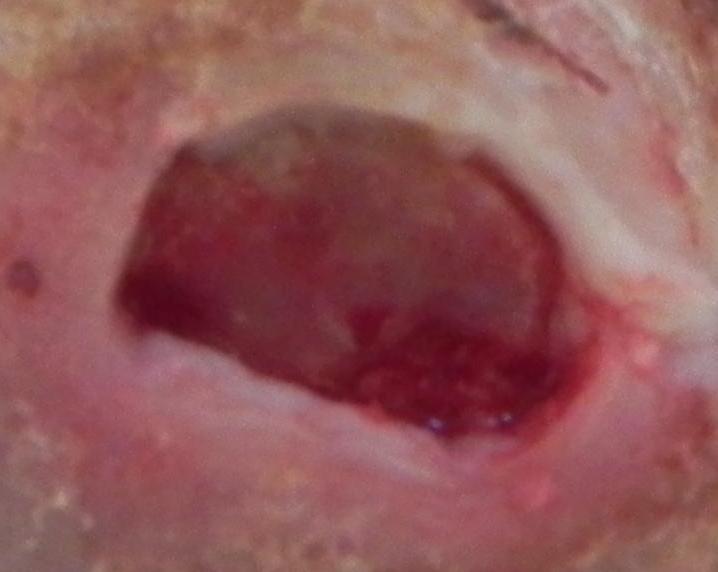}& 
		\includegraphics[width=3cm,height=3cm]{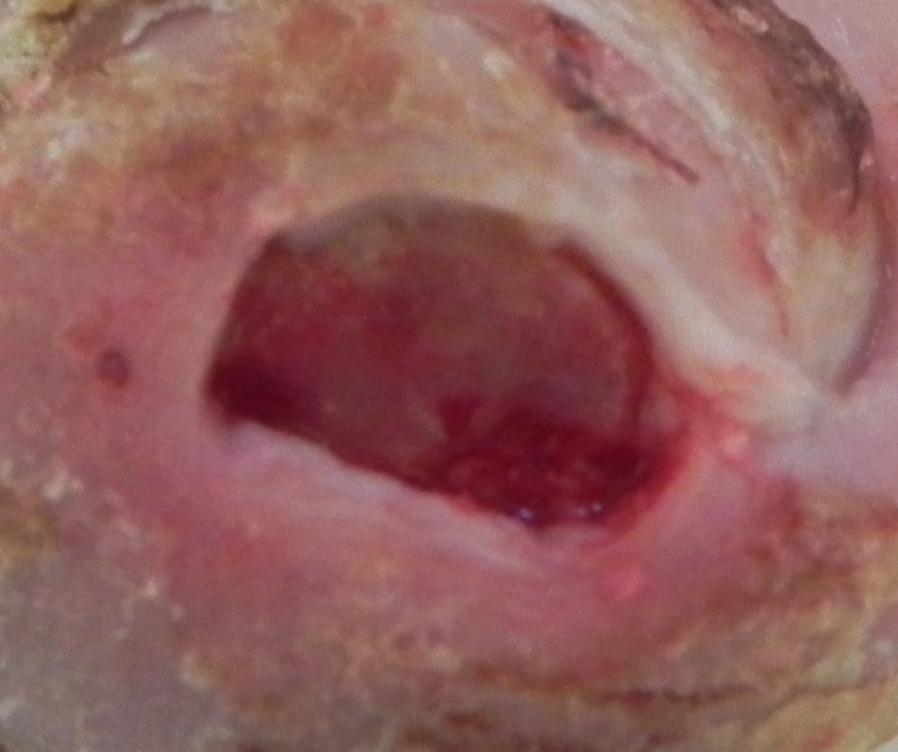}&    
		\\
		
		(a) Image & (b) Ist MAG & (c) 2nd MAG & (d) 3rd MAG &
		\\
	\end{tabular}     
	
	\caption{Natural Data-augmentation produced from the original image with different magnifications (three magnifications in this experiment). MAG refers to magnification}
	\label{fig:resultsVisual223}
\end{figure*}

\begin{figure*}[!t]
	\centering
	\small
	\begin{tabular}{cccc}
		\includegraphics[width=3cm,height=3cm]{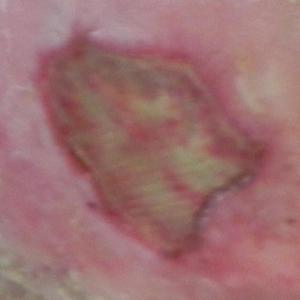} &
		\includegraphics[width=3cm,height=3cm]{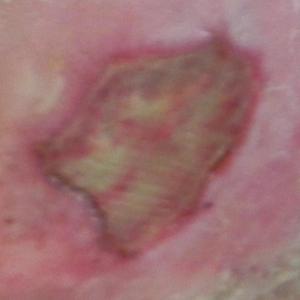} &
		\includegraphics[width=3cm,height=3cm]{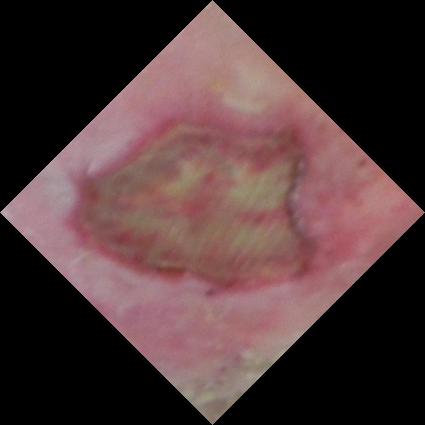}&
		\includegraphics[width=3cm,height=3cm]{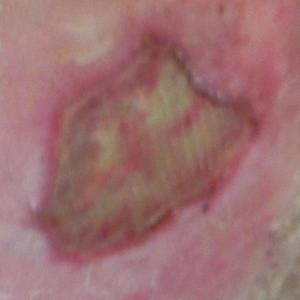}\\
		(a) Image & (b) Mirror& (c) 45$^{\circ}$ & (d) 90$^{\circ}$\\
		\includegraphics[width=3cm,height=3cm]{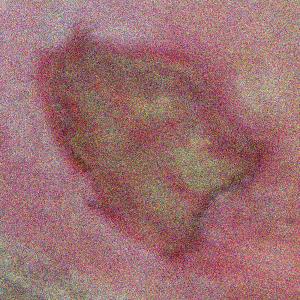}&
		\includegraphics[width=3cm,height=3cm]{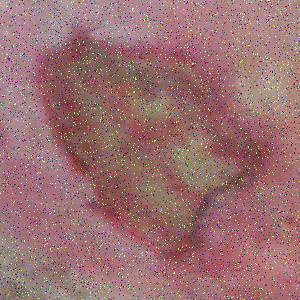}&
		\includegraphics[width=3cm,height=3cm]{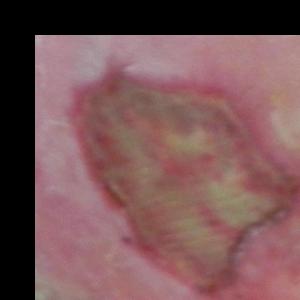}&
		\includegraphics[width=3cm,height=3cm]{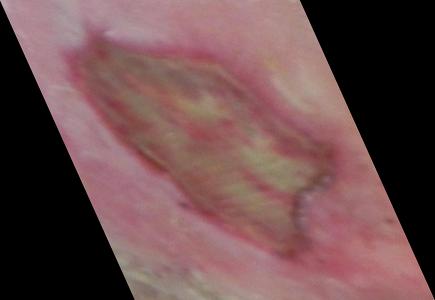}\\
		(e) Gaussian Noise & (f) Salt and pepper & (g)Translate & (h)Shear
		\\
	\end{tabular}     
	
	\caption{After magnification, different types of data-augmentation is achieved by the proposed Natural Data-augmentation}
	\label{fig:Aug}
\end{figure*}

In the DFU dataset, the images (size )varies between 1600 $\times$ 1200 and 3648 $\times$ 2736) depending on the cameras used to capture the data. In deep learning, data augmentation is envisioned as an important tool to improve the performance of algorithms. As shown in Fig. \ref{fig:AreaDFUs}, approximately 92\% of DFU cases have area between 0\% to  20\% on foot images. In common data-augmentation, the number of techniques used such as flip, rotation, random scale, random crop, translation, and Gaussian noise to perform augment in the dataset. Since DFU occupies a very small percentage of the total area of foot images, there is a risk of missing the region of interests by using important augmentation technique such as random scale, crop, and translation. Hence, Natural Data-augmentation is more suitable for the DFU evaluation rather than common data-augmentation. This augmentation technique helps in assisting the machine algorithms to pinpoint ROI of DFU on foot images and focus on finding the strong features that exists in this area. We used the deep learning-based localization method, Faster-RCNN with InceptionResNetV2, to get ROI of the DFU on foot images \cite{goyal2018region, huang2016speed}.  Depending upon the size of DFU and image, the natural data-augmentation on the DFU dataset with different magnification is demonstrated in Fig. \ref{fig:resultsVisual223}. Flexible parameters can be used to choose the number of magnification factors (3 in this classification), as well as magnification distance, which can be adjusted from a single DFU image by natural augmentation. After magnification, further, data-augmentation is achieved with the help of angles, mirror, gaussian noise, contrast, sharpen, translation, shearing using our proposed methods as shown in Fig. \ref{fig:Aug}. 

\begin{figure*}
	\centering
	\includegraphics[width=1\textwidth]{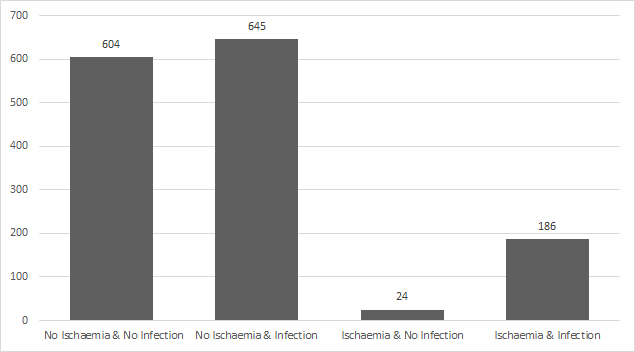}
	\caption{Distribution of ischaemia and infection cases as multi-class classification problem.}
	\label{fig:Dist}
\end{figure*}

As shown in Table \ref{tab:sinbad2}, the number of DFU patches generated by cropping multiple DFU on foot images and augmented patches are generated by natural data-augmentation (Fig. \ref{fig:resultsVisual223}) and  different data augmentations (Fig. \ref{fig:Aug}). The total number of cases for ischaemia and non-ischaemia in this DFU dataset is imbalanced (1249 cases vs 210 cases) whereas infection (628 cases) and non-infection (831 cases) are fairly balanced as shown in Table \ref{tab:sinbad2}.  We performed binary classification of ischaemia and infection with machine learning algorithms because for multi-class classification, this DFU dataset is imbalanced especially for cases (Ischaemia and No Infection) as shown in \ref{fig:Dist}.

\begin{table}[]
	\centering
	\addtolength{\tabcolsep}{-2pt}
	\renewcommand{\arraystretch}{1.5}
	\caption{The number of Infection and ischaemia cases, number of DFU patches and augmented patches using Natural Data-augmentation in DFU Dataset}
	\label{tab:sinbad2}
	\scalebox{0.8}{
		\begin{tabular}{llccc}
			\hline
			Category                             & Definition                                           & Cases & DFU patches  & Augmented patches\\\hline\hline
			\multirow{2}{*}{Ischaemia}            & Absent & 1249   & 1431& 4935     \\
			& Present       & 210      &235 & 4935      \\\hline
			Total images             &   &  1459 &1666&9870 \\\hline
			\multirow{2}{*}{Bacterial infection} & None                                                 &   628& 684& 2946        \\
			& Present                                              & 831  &982& 2946          \\\hline
			Total images                                                                     & &1459 &1666  &  5892 \\ \hline       
	\end{tabular}}
\end{table}  

\subsection{Handcrafted Superpixel Color Descriptors}
We investigated the use of human design features with traditional machine learning on the binary classification of infection and ischaemia. Our first attempt was experimenting with texture descriptors (Local Binary Patterns and Histogram of Gradient) and color descriptors as used in related works \cite{goyal2018dfunet, goyal2018robust}. However, we achieved very poor results for these binary classification problems. Hence, we propose a novel Superpixel Color Descriptors (SPCD) to extract the colors region of interest from DFU images that could be the important visual cues for the identification of ischaemia and infection in DFU. In the first step, we used a SLIC superpixels technique to produce superpixel over-segmentation of DFU patches based on pixel color and intensity values \cite{achanta2010slic}. SLIC superpixels technique performs a localized \textit{k}-means optimization in the 5-D CIELAB color and image space to cluster pixels as described by equations \ref{eq:1} - \ref{eq:4}:

\begin{equation} \label{eq:1}
S=\sqrt{\frac{N}{k}}
\end{equation}

\begin{equation} \label{eq:2}
D_{s} =d_{l a b}+\frac{m}{S} d_{x y} 
\end{equation}

\begin{equation} \label{eq:3}
d_{lab}=\sqrt{\left(l_{k}-l_{i}\right)^{2}+\left(a_{k}-a_{i}\right)^{2}+\left(b_{k}-b_{i}\right)^{2}}
\end{equation}

\begin{equation} \label{eq:4}
d_{x y} =\sqrt{(x_{k}-x_{i})^{2}+(y_{k}-y_{i})^{2}} 
\end{equation}

where in eq. \ref{eq:1}, \textit{S} is the approximate size of a superpixel, \textit{N} is the number of pixels and \textit{k}  is the number of superpixels; 
in eq. \ref{eq:2}, \textit{D\textsubscript{s}} is the sum of the lab distance (d\textsubscript{lab})and the xy plane distance (d\textsubscript{xy}); in eq. \ref{eq:3}, \textit{l}, \textit{a} and \textit{b} represent the lab colorspace; and in eq. \ref{eq:4}, \textit{x} and \textit{y} represent the pixel positions.

In the second step, the mean RGB color value of each superpixel is computed and applied to each superpixel (\textit{S}) denoted by: 

\begin{equation}\label{eq:5}
S_i = mean(P(R,G,B)) , i = 1,\ldots, k
\end{equation}

where in eq. \ref{eq:5}, \textit{P(R,G,B)} is the pixel values of R,G,B channel in each \textit{ith} position of \textit{S} and \textit{k} is total number of superpixels in the image. 

\begin{figure*}
	\centering
	\includegraphics[width=1\textwidth]{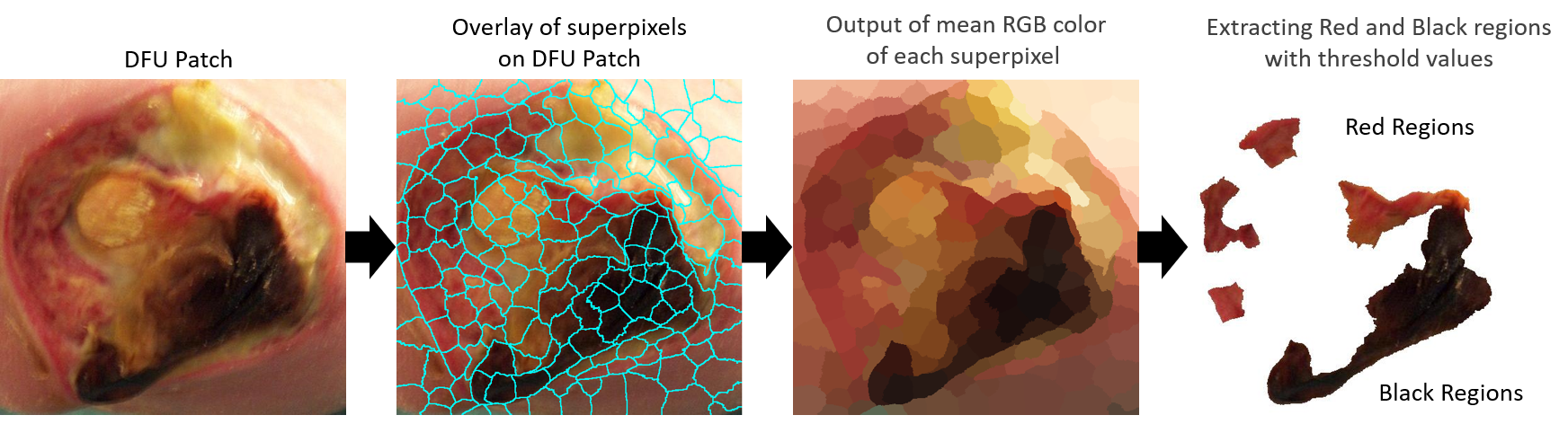}
	\caption{Example of extracting red and black regions from DFU patch with proposed Superpixel Color Descriptor algorithm which was then used to inform identification of ischaemia and infection. The k value of 200 for superpixel algorithm effectively oversegmented the DFU patches.}
	\label{fig:BR}
\end{figure*}

Finally, with a different number of superpixels and threshold values from each color channel, we extracted regions of two particular colors of interest that are red and black from the DFU patches. For these classification tasks, we used the number of superpixels (k=200) and threshold values (T1: 0.40,0.45,0.50,.055,0.60; T2: 0.15,0.20,0.25,0.30,0.35) to extract the color features from DFU patches of 256$\times$256. The threshold values are used to restrict the intensities of red and black pixels to be utilized as handcrafted features. Hence, we utilised a feature vector of 10 with SPCD algorithm along with texture descriptors (LBP, HOG) and color features (RGB, CIELAB) to train traditional machine learning approaches. The pseudocode for the SPCD algorithm is explained in Algorithm \ref{euclid}. The example of extracting color features using our novel SPCD algorithm is shown in Fig. \ref{fig:BR}.

\begin{algorithm}
	\caption{Pseudocode for the Superpixel Color Descriptors Extraction}\label{euclid}
	\begin{algorithmic}[1]
		\State Over-segmentation of DFU patch with SLIC superpixel is performed;
		\State Mean RGB value of each superpixel is calculated and applied;
		\State Initialize variable S\_Red \ \& \ S\_Black to 0
		\Procedure{RedAndBlackRegion}{}
		\For{$ each \ Superpixel(S\textsubscript{i})$}
		\If {$S\textsubscript{i}(R) > T\textsubscript{1}*(S\textsubscript{i}(R)+S\textsubscript{i}(G)+S\textsubscript{i}(B))$} \Return S\_Red= S\_Red + 1
		\EndIf
		\If {$S\textsubscript{i}(R)< T\textsubscript{2} \ \& \ S\textsubscript{i}(G)< T\textsubscript{2} \ \& \ S\textsubscript{i}(B)< T\textsubscript{2}$}
		\Return S\_Black= S\_Black + 1
		\EndIf
		\State $RedColorFeature= \ S\_Red \ \div \  n$
		\State $BlackColorFeature= \ S\_Black \ \div \  n$
		\EndFor
		\EndProcedure
	\end{algorithmic}
\end{algorithm}

For these classification problems, we experimented with a number of classifiers with standard hyper-parameters on these color features. BayesNet, Random Forest, and Multilayer Perceptron were selected and achieved the highest accuracy among other machine learning classifiers.

\subsection{Deep Learning Approaches}
For comparison with the traditional features, deep learning algorithms are used to perform binary classification to classify (1) infection and non-infection; and (2) ischaemia and non-ischaemia classes in DFU patches.  For this work, we fine-tune (transfer learning from pre-trained models) the CNN models, i.e. Inception-V3, ResNet50, and InceptionResNetV2 \cite{szegedy2016rethinking, szegedy2016inception, he2016deep}. To train the CNN networks, we froze the weights of the first few layers of the pre-trained networks for common features, such as edges and curves. Subsequently, layers of networks are unfrozen to focus on learning dataset-specific features.

\begin{figure*}
	\centering
	\includegraphics[width=1\textwidth]{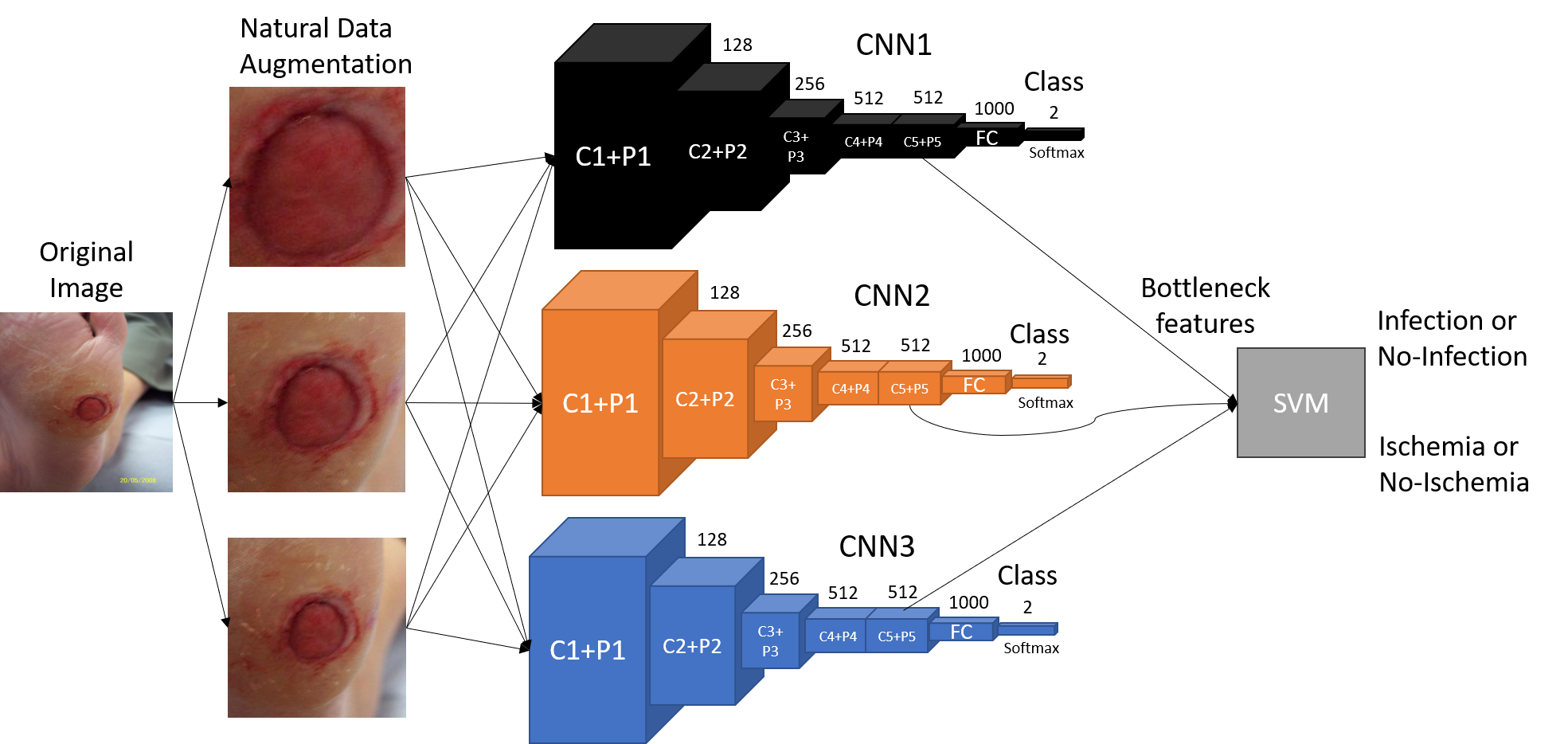}
	\caption{Extracting bottleneck features from CNNs and fed into SVM classifier to perform binary classification of ischaemia and infection, where C1-C5 are convolutional layers, P1-P5 are pooling layers and FC is fully connected layer. \textbf{Note: The CNNs in this figure are just representations of general CNNs architecture and do not represent the original CNN architectures of Inception-V3, ResNet50, and InceptionResNetV2.}}
	\label{fig:ECNN}
\end{figure*}

Additionally, we utilized the Ensemble CNN method, which is a very effective CNN approach to obtain very good accuracy on difficult datasets.  The Ensemble CNN model combines the bottleneck features from multiple CNN models (Inception-V3, ResNet50, and InceptionResNetV2), and use SVM classifier to produce predictions, as shown in Fig. \ref{fig:ECNN}. 

\section{Results and Discussion}\label{sec:7.5}
Both infection and ischaemia datasets were split into 70\% training, 10\% validation and 20\% testing sets and we adopted the 5-fold cross-validation technique. We utilized the natural data-augmentation technique for training and validation sets in both traditional machine learning and deep learning approaches. Hence, in this ischaemia dataset, we used approximately 11,564 patches, 1,652 patches, and 3,304 patches in training, validation, and testing sets respectively whereas, in the infection dataset, we used 7,136 patches (training), 1,019 patches (validation), and 2,038 patches (testing) from the 2611 original foot images. As mentioned previously, we used both handcrafted traditional machine learning (henceforth TML) models and CNN models to perform the classification task and utilized 256$\times$256 RGB images as input for TML and InceptionV3, AlexNet, and ResNet50. For InceptionResNetV2, we resized the dataset to 299$\times$299. For this experiment, TensorFlow is used for deep learning and Matlab is used for traditional machine learning approaches.

In Table \ref{tab:tradmachineIsc} and  \ref{tab:tradmachineInf}, we report \textit{Accuracy}, \textit{Sensitivity}, \textit{Precision}, \textit{Specificity}, \textit{F-Measure}, \textit{Matthew Correlation Coefficient (MCC)} and \textit{Area under the ROC curve (AUC)} as our evaluation metrics. 

%In medical imaging, \textit{Sensitivity} and \textit{Specificity} are considered reliable evaluation metrics for classifier completeness. 

\begin{table*}[]
	\centering
	\addtolength{\tabcolsep}{-2pt}
	\renewcommand{\arraystretch}{2}
	\caption{The performance measures of binary classification of ischaemia by our proposed handcrafted traditional machine learning and CNN approaches.}
	\label{tab:tradmachineIsc}
	\scalebox{0.58}{
		\begin{tabular}{lccccccc}
			\hline
			& \textit{Accuracy} & \textit{Sensitivity} & \textit{Precision} & \textit{Specificity} & \textit{F-Measure} & \textit{MCC Score} & \textit{AUC Score} \\ \hline\hline
			BayesNet                             & 0.785$\pm$0.022       & 0.774$\pm$0.034       & 0.809$\pm$0.034     & 0.800$\pm$0.027    & 0.790$\pm$0.020     & 0.572$\pm$0.044 & 0.783   \\
			Random Forest               & 0.780$\pm$0.041       & 0.739$\pm$0.049       & 0.872$\pm$0.029     & 0.842$\pm$0.034    & 0.799$\pm$0.033     & 0.571$\pm$0.078  & 0.780 \\
			Multilayer Perceptron               &0.804$\pm$0.022       & 0.817$\pm$0.040     & 0.787$\pm$0.046    & 0.795$\pm$0.031     & 0.800$\pm$0.023 & 0.610$\pm$0.045    & 0.804 \\
			InceptionV3 (CNN)    & 0.841$\pm$0.017       & 0.784$\pm$0.045       & 0.886$\pm$0.018    & 0.898$\pm$0.022    & 0.831$\pm$0.021    & 0.688$\pm$0.031   & 0.840\\
			ResNet50 (CNN)   & 0.862$\pm$0.018       & 0.797$\pm$0.043       & 0.917$\pm$0.015     & 0.927$\pm$0.017   & 0.852$\pm$0.022     & 0.732$\pm$0.032   & 0.865\\
			InceptionResNetV2 (CNN)   & 0.853$\pm$0.021       & 0.789$\pm$0.054       & 0.906$\pm$0.017     & 0.917$\pm$0.019    & 0.842$\pm$0.027     & 0.714$\pm$0.039 & 0.851  \\
			Ensemble (CNN)   & \textbf{0.903$\pm$0.012}       & \textbf{0.886$\pm$0.035}       & \textbf{0.918$\pm$0.019}     & \textbf{0.921$\pm$0.021}    & \textbf{0.902$\pm$0.014}     & \textbf{0.807$\pm$0.022} & \textbf{0.904} \\ \hline
	\end{tabular}}
\end{table*}
\begin{table*}[]
	\centering
	\addtolength{\tabcolsep}{-2pt}
	\renewcommand{\arraystretch}{2}
	\caption{The performance measures of binary classification of Infection by our proposed handcrafted traditional machine learning and CNN approaches.}
	\label{tab:tradmachineInf}
	\scalebox{0.58}{
		\begin{tabular}{lccccccc}
			\hline
			& \textit{Accuracy} & \textit{Sensitivity} & \textit{Precision} & \textit{Specificity} & \textit{F-Measure} & \textit{MCC Score} & \textit{AUC Score}\\ \hline\hline
			BayesNet                             & 0.639$\pm$0.036       & 0.619$\pm$0.018       & 0.653$\pm$0.039     & 0.660$\pm$0.015    & 0.622$\pm$0.079     & 0.290$\pm$0.070 & 0.643   \\
			Random Forest              & 0.605$\pm$0.025       & 0.608$\pm$0.025       & 0.607$\pm$0.037     & 0.601$\pm$0.069    & 0.606$\pm$0.012     & 0.211$\pm$0.051   &0.601\\
			Multilayer Perceptron              &0.621$\pm$0.026       & 0.680$\pm$0.023     & 0.622$\pm$0.057    & 0.570$\pm$0.023     & 0.627$\pm$0.074 & 0.281$\pm$0.055     &0.619\\
			InceptionV3 (CNN)    & 0.662$\pm$0.014       & 0.693$\pm$0.038       & 0.653$\pm$0.015    & 0.631$\pm$0.034    & 0.672$\pm$0.019    & 0.325$\pm$0.029  & 0.662 \\
			ResNet50 (CNN)   & 0.673$\pm$0.013       & 0.692$\pm$0.051       & 0.668$\pm$0.023     & 0.654$\pm$0.051    & 0.679$\pm$0.019    & 0.348$\pm$0.028   &0.673\\
			InceptionResNetV2 (CNN)   & 0.676$\pm$0.015      & 0.688$\pm$0.052       & 0.672$\pm$0.015     & 0.664$\pm$0.039    & 0.680$\pm$0.024     & 0.352$\pm$0.031 &0.678 \\ 
			Ensemble (CNN)   & \textbf{0.727$\pm$0.025}       & \textbf{0.709$\pm$0.044}       & \textbf{0.735$\pm$0.036}     & \textbf{0.744$\pm$0.050}    & \textbf{0.722$\pm$0.028}     & \textbf{0.454$\pm$0.052}  & \textbf{0.731}\\ \hline\hline
	\end{tabular}}
\end{table*}

\begin{figure}
	\centering
	\includegraphics[width=0.8\textwidth]{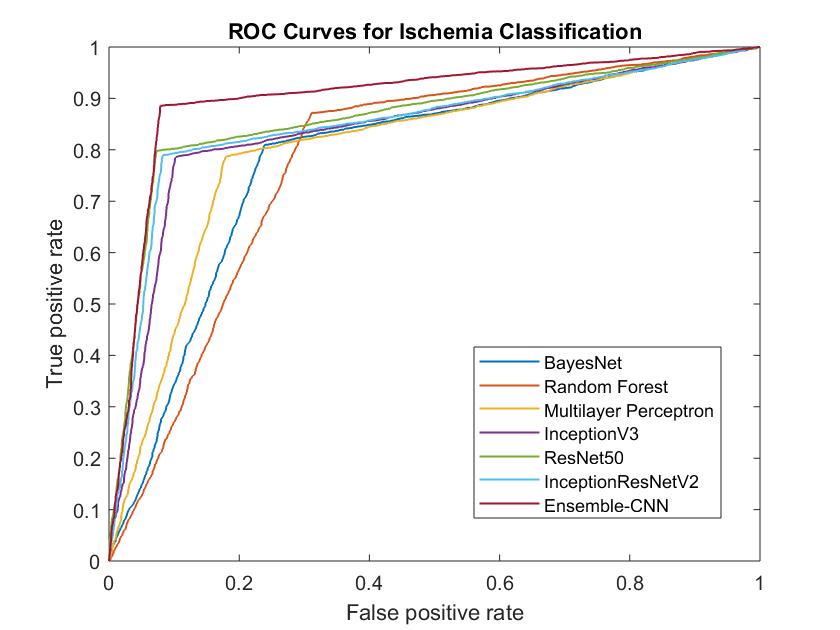}
	\caption{ROC curve for all TML and CNN methods for ischaemia classification.}
	\label{fig:ROCischaemia}
\end{figure}

\begin{figure}
	\centering
	\includegraphics[width=0.8\textwidth]{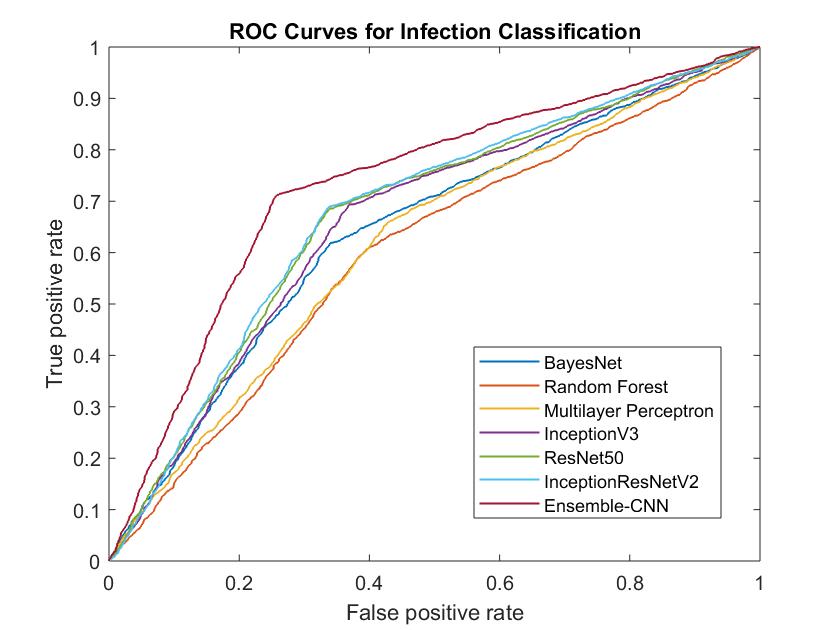}
	\caption{ROC curve for all TML and CNN methods for Infection classification.}
	\label{fig:ROCInfection}
\end{figure}

When comparing the performance of the computerized methods and our proposed techniques, CNNs performed better in the binary classification of ischaemia than infection despite more imbalanced data in the ischaemia dataset, due to more cases of non-ischaemia in the dataset. The average performance of all the models in terms of accuracy in the ischaemia dataset was 83.3\% which is notably better than the average accuracy of 65.8\% in infection dataset. Similarly, \textit{MCC Score} and \textit{AUC Score} are considered to be viable performance measures to compare the classification results. We obtained an average \textit{MCC Score} and \textit{AUC Score} for ischaemia classification of 67.1\% and 83.2\% respectively, as compared to the infection classification of 32.3\% and 65.8\% respectively. The ROC curves for all the algorithms, including TML and CNNs for binary classification of ischaemia and infection, are shown in Fig. \ref{fig:ROCischaemia} and \ref{fig:ROCInfection}. When comparing the performances in ischaemia classification of TML and CNNs, CNNs (86.5\%) performed better than the TML models (79\%). Similarly, in infection classification, the accuracy of CNNs (68.4\%) performed better than TML (62.1\%) with a margin of 6.3\%. Notably, Ensemble CNN method achieved the highest score in all performance measures in both ischaemia and infection classification.

\textit{Sensitivity} and \textit{Specificity} are considered important performance measures in medical imaging. The ensemble method yielded high \textit{Sensitivity} for the ischaemia dataset with a margin of 6.9\% from the second best performing algorithm multilayer perceptron. Interestingly, a multilayer perceptron performed worst in the \textit{Specificity} with a score of 79.5\%. For \textit{Specificity} in the ischaemia dataset, the ensemble method again obtained the highest score of 92.9\% which is marginally better than ResNet50 (92.7\%).

In infection classification, both TML and CNN methods received moderate scores in the performance measures. Again, CNN methods performed better than TML methods achieving the highest score in all performance measures. The Ensemble CNN method performed better than other CNN classifiers especially for \textit{Specificity} with a score of  74.4\% in infection classification with a notable margin of 8\% than the second-best performing algorithm InceptionResNetV2(66.4\%). For \textit{Sensitivity}, all the CNNs performed marginally well with Ensemble method achieving the highest score of 70.9\%. When comparing the performance of TML methods, Multilayer Perceptron (68.0\%) performed well in \textit{Sensitivity}, whereas BayesNet (66\%) better in \textit{Specificity}.  

\begin{figure}[!t]
	\centering
	\small
	\begin{tabular}{ccccc}
		\includegraphics[width=2.85cm,height=2.85cm]{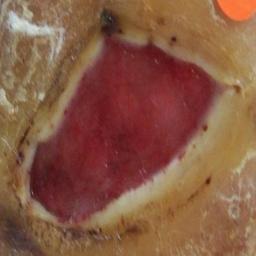} &
		\includegraphics[width=2.85cm,height=2.85cm]{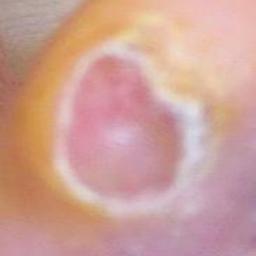}&
		\includegraphics[width=2.85cm,height=2.85cm]{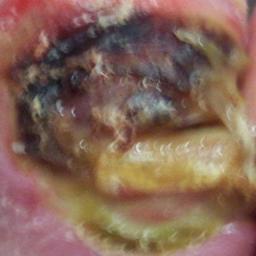}& 
		\includegraphics[width=2.85cm,height=2.85cm]{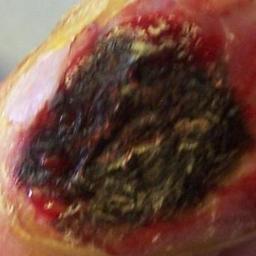}\\  
		(a) & (b)  & (c)& (d)&
		\\
		\multicolumn{2}{l}{Accurate non-ischaemia cases}  & \multicolumn{2}{l}{Accurate ischaemia cases}&
		\\
	\end{tabular}     
	
	\caption{Examples of correctly classified cases by Ensemble-CNN on ischaemia dataset. (a) and (b) represent non-ischaemia cases. (c) and (d) represent ischaemia cases.  }
	\label{fig:misclass1}
\end{figure}

\begin{figure}[!t]
	\centering
	\small
	\begin{tabular}{ccccc}
		\includegraphics[width=2.85cm,height=2.85cm]{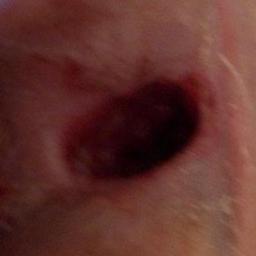} &
		\includegraphics[width=2.85cm,height=2.85cm]{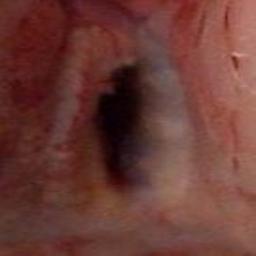}&
		\includegraphics[width=2.85cm,height=2.85cm]{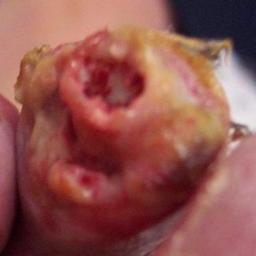}& 
		\includegraphics[width=2.85cm,height=2.85cm]{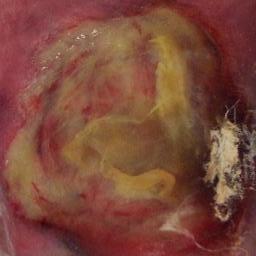}\\  
		(a) & (b)  & (c)& (d)&
		\\
		
		\multicolumn{2}{l}{Misclassified non-ischaemia cases}  & \multicolumn{2}{l}{Misclassified ischaemia cases}&
		\\
	\end{tabular}     
	
	\caption{Examples of misclassified cases by Ensemble-CNN on ischaemia dataset. (a) and (b) represents non-ischaemia cases. (c) and (d) represents ischaemia cases.  }
	\label{fig:misclass2}
\end{figure} 

\begin{figure}[!t]
	\centering
	\small
	\begin{tabular}{ccccc}
		\includegraphics[width=2.85cm,height=2.85cm]{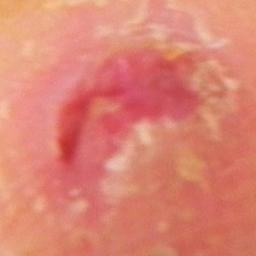} &
		\includegraphics[width=2.85cm,height=2.85cm]{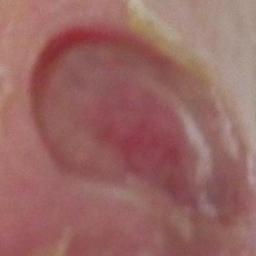}&
		\includegraphics[width=2.85cm,height=2.85cm]{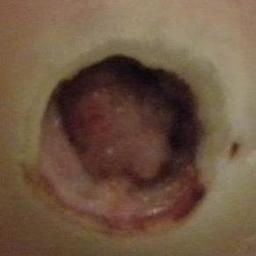}& 
		\includegraphics[width=2.85cm,height=2.85cm]{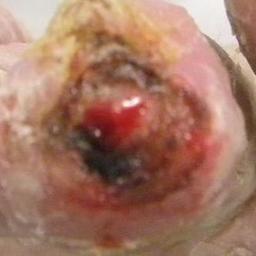}\\  
		
		(a) & (b)  & (c)& (d)&
		\\
		\multicolumn{2}{l}{Accurate non-infection cases}  & \multicolumn{2}{l}{Accurate infection cases}&
		\\
	\end{tabular}     
	
	\caption{Examples of correctly classified cases by Ensemble-CNN on Infection dataset. (a) and (b) represents non-infection cases. (c) and (d) represents infection cases.  }
	\label{fig:misclass3}
\end{figure}

\begin{figure}[!t]
	\centering
	\small
	\begin{tabular}{ccccc}
		\includegraphics[width=2.85cm,height=2.85cm]{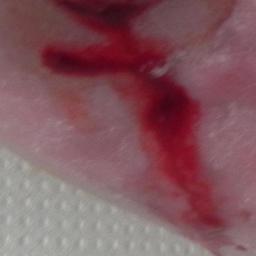} &
		\includegraphics[width=2.85cm,height=2.85cm]{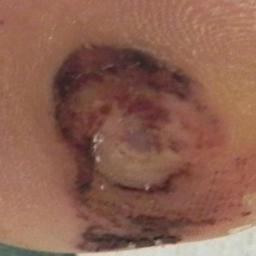}&
		\includegraphics[width=2.85cm,height=2.85cm]{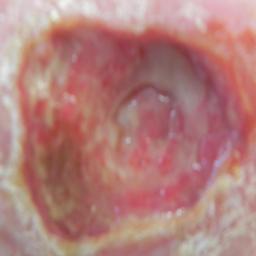}& 
		\includegraphics[width=2.85cm,height=2.85cm]{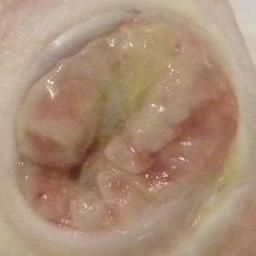}\\  
		
		(a) & (b)  & (c)& (d)&
		\\
		\multicolumn{2}{l}{Misclassified non-infection}  & \multicolumn{2}{l}{Misclassified infection cases}&
		\\
	\end{tabular}     
	
	\caption{Examples of misclassified cases by Ensemble-CNN on Infection dataset. (a) and (b) represents non-infection cases. (c) and (d) represents infection cases.  }
	\label{fig:misclass4}
\end{figure}

\subsection{Experimental Analysis and Discussion}
Assessment of DFU with computerized methods is very important for supporting global healthcare systems through improving triage and monitoring procedures and reducing hospital time for patients and clinicians. This preliminary experiment is focused on automatically identifying the important conditions of ischaemia and infection of DFU. The main aim of this experiment was to identify ischaemia and infection from images of the feet using machine learning. We have illustrated examples of correctly and incorrectly classified cases in both binary classifications of ischaemia (Fig. \ref{fig:misclass1} and \ref{fig:misclass2}) and infection (Fig.  \ref{fig:misclass3} and \ref{fig:misclass4}). As for the misclassified cases, there are huge intra-class dissimilarities and inter-class similarities between (1) infection and non-infection; (2) ischaemia and non-ischaemia cases in the DFU that make classifiers difficult to predict the correct class. Additionally, there are other influencing factors in the classification of these conditions such as lighting conditions, marks and skin tone. In misclassified cases of non-ischaemia as shown in Fig. \ref{fig:misclass2}, the cases (a) and (b) are hindered by the lighting condition (shadow) respectively, whereas in the (c) and (d) misclassified ischaemia cases, the ischaemia features may be too subtle to be recognised from the images by the algorithm. Alternatively it is likely we needed a more sensitive objective measure of the ground truth from vascular assessments. We found that shadows are particularly problematic because machine learning algorithms can be deceived by shadows especially in determining the important conditions such as ischaemia. In Fig. \ref{fig:misclass4}, misclassified cases of non-infection, the presence of blood in the case (a), whilst case (b) belongs to one of the rare cases with the presence of ischaemia and non-infection. In misclassified infection cases, the visual indicators of infection were likely too subtle, or we needed more sensitive objective ground truth provided through blood analysis. 

In this work, we used the proposed natural data-augmentation with the help of DFU localisation to create DFU patches from full-size foot images. These patches are useful to focus more on finding the visual indicators for important factors of DFU such as infection and ischaemia. Then, we investigated the use of both TML and CNNs to determine these conditions as binary classification. In this experiment, we received very good performance in terms of correctly classifying ischaemia despite the imbalanced cases in the DFU dataset. However, in the case of infection, the classifiers did not perform as well, since the condition of infection is hard to recognise from the foot images even by experienced medical experts specialized in DFU and therefore likely requires ground truth determined using objective blood tests to identify bacterial infection.

Current research focuses on ischaemia and infection recognition in medical classification systems, which requiring the guidance of medical experts specialized in DFU. To develop a computer-aided tool for medical experts in remote foot analysis, i.e. a remote DFU diagnosis system, the following are challenges need to be addressed:

\begin{enumerate}
	\item Recognition of the ischaemia and infection with machine learning algorithms as an important proof-of-concept study for foot pathologies classification. Further analysis of each pathology on foot images is required according to the medical classification systems, such as the University of Texas Classification of DFU \cite{lavery1996classification} and SINBAD Classification System \cite{ince2008use}. This requires close collaboration with medical experts specialized in DFU.  
	\item Deep learning algorithms need substantial datasets to obtain very good accuracy, especially for medical imaging. This experiment included an imbalanced DFU dataset (1459 foot images) for both ischaemia and infection conditions. In the future, if these algorithms were to train with a larger number of a more balanced dataset, it can possibly improve the recognition of ischaemia and infection.
	\item A study of the performance of algorithms on different types of capturing devices is an important aspect of future work. This experiment evaluates the performance of machine learning algorithms on the DFU dataset collected with different cameras (heterogeneous sources of data). This leads to more variability of image characteristics. Since the algorithms have to deal with more heterogeneous patterns and characteristics that are not intrinsic to the pathology itself. In this experiment, we know that three types of devices were used, we do not have the information on the association of images and the type of devices.
	\item The current ground truth is based on visual inspection by experts only and not supported by the medical notes or clinical tests (vascular assessment for ischaemia and blood tests to identify the presence of any bacterial infection). Furthermore, DFU images were debrided before these images were captured. Hence, the debridement of DFU removes important visual indicators of infection such as colored exudate. Therefore, the sensitivity and specificity of these algorithms could be further improved in the future, by feeding in ground truth from clinical tests such as vascular assessments (ischaemia) and blood tests (to identify the presence of any bacterial infection).
	\item Current clinical practice obtains the foot photo using different camera models, poses and illumination. It is a great challenge for a computer algorithm to predict the depth and the size of the wound based on non-standardized images. Standardized dataset, such as the data collection method proposed by Yap et al. \cite{yap2018new} will help to increase the accuracy of the DFU diagnosis system.
	\item  Dataset annotation is a laborious process, particularly for medical experts to label the foot pathologies into 16 classes according to the University of Texas classification system. To reduce the burden upon medical experts in the delineation and annotation of the dataset, there is an urgent need to focus on developing unsupervised or self-supervised machine learning techniques. 
	\item Collecting the time-line dataset is crucial for early detection of key pathologies. This will enable monitoring of foot health and changes longitudinally, where medical experts and computer algorithms can learn the early signs of DFU. In the longer-term, the DFU diagnosis system will be able to predict the healing process of ulcers and prevent DFU before it happens.
	\item A smart-phone app could be developed for remote triage and monitoring of DFU. To scale-up the DFU diagnosis system, the application should run on multiple devices, irrespective of the platform and/or the operating system.
\end{enumerate}

\section{Conclusion}
In this work, we trained various classifiers based on traditional machine learning algorithms and CNNs to discriminate the conditions of: (1) ischaemia and non-ischaemia; and (2) infection and non-infection related to a given DFU. We found high-performance measures in the binary classification of ischaemia, compared to moderate performance by classifiers in the classification of infection. It is vital to understand the features of both conditions in relation to the DFU (ischaemia and infection) from a computer vision perspective. Determining these conditions especially infection from the non-standard foot images is very challenging due to: (1) high visual intra-class dissimilarities and inter-class similarities between classes;
(2) the visual indicators of infection and ischaemia potentially being too subtle in DFU; (3) objective medical tests for vascular supply and bacterial infection are needed to provide more objective ground truth and further improve the classification of these conditions; and (4) other factors such as lighting conditions, marks and skin tone are important to incorporate into the prediction. 

With a more balanced dataset and improved data capturing of DFU, the performance of these methods could be improved in the future. Further optimization in hyper-parameters of both deep learning and traditional machine learning methods could improve the performance of algorithms on this dataset. Ground truths enhanced by clinical tests for the ischaemia and infection may provide further insight and further improvement of algorithms even where there is no apparent visual indicator by eye. In the case of infection even after debridement, ground truth informed by blood tests for infection may yield improvements to sensitivity and specificity even in the absence of overtly obvious visual indicators. This work has the potential for technology that may transform the recognition and treatment of diabetic foot ulcers and lead to a paradigm shift in the clinical care of the diabetic foot. 

\section*{Acknowledgements}

The authors express their gratitude to Lancashire Teaching Hospitals and the clinical experts for their extensive support and contribution in carrying out this research. We would like to thank Kim's English Corner (https://kimsenglishcorner.com) for proofreading. 	


\section*{Reference}
 \bibliographystyle{elsarticle-num} 
 \bibliography{Ref}


%\begin{thebibliography}{00}


%\bibitem{}

%\end{thebibliography}
\end{document}